\documentstyle[12pt]{article}
\begin{document}
\begin{center}
{\bf  $\mbox{SU(3)}_L\otimes \mbox{U(1)}_N$ MODEL FOR  RIGHT-HANDED\\
 NEUTRINO NEUTRAL CURRENTS }\\
\vspace{2cm}
{\bf Hoang Ngoc Long}\\
{\it Institute of Theoretical Physics,
National Centre for Natural Science and
Technology,\\
P.O.Box 429, Bo Ho, Hanoi 10000, Vietnam.}\\
\vspace{1cm}

{\bf Abstract}\\
\end{center}
A model based on the $\mbox{SU(3)}_L\otimes \mbox{U(1)}_N$
gauge group, in which neutrinos have  right-handed neutral
currents is considered. We argue that in order to have a result
consistent with low-energy one, the  right-handed neutrino
component must be treated as correction instead of an
equivalent spin state.

PACS number(s): 12.15.Mm, 12.15.Ff, 12.15.Cc\\
\vspace{5cm}

\begin {center}
( To appear in {\it Physical Review D} )
\end{center}

\newpage
\noindent

For a long time, neutrino is one of the most attracted
subjects in particle physics. It is well known that in
the standard model (SM) neutrinos have only left-handed (LH)
currents. Whether neutrinos have right-handed (RH) currents
is still an unresolved question.

Recently, RH neutrino currents have been included in
a model based on the $SU(3)_C\otimes SU(3)_L \otimes U(1)_N$
gauge group~\cite{mpp}.  However, by some reasons,
consequences of this model (model I as called in~\cite{mpp})
are too different from those of the SM. For example,
{\it the magnitude of the neutral couplings of the right-handed
neutrinos coincides with that of the left-handed neutrinos
in the SM} (see Eqs.(44,51) in~\cite{mpp}). If so, in order to
get results consistent with low energy phenomenology such as
$\nu_\mu-e$ and $\bar{\nu}_\mu-e$ scatterings, $g_L(e)$ must be
replaced by $g_R(e)$ and vice versa~\cite{bp}. It is obvious
that this statement is unacceptable.

It is to be noted that a similar model has been proposed
in~\cite{flt} (for details see~\cite{hnl}) in which RH neutrinos
by opposition to our model, give contribution to neutral
currents of the LH  neutrinos.
The purpose of  this report is to briefly present the model
(for more details in this model, the reader can find
in~\cite{hepph}).

Our  model deals with nine leptons and nine quarks. There are
three left- and right-handed neutrinos ($\nu_e, \nu_\mu ,\nu_\tau$),
three charged leptons ($e, \mu, \tau$), four quarks with charge 2/3,
and five quarks with charge -1/3. Fermion content of this model
is almost the same as in~\cite{hnl} with the unique change in a
bottom of the lepton triplets
\begin{equation}
f^{a}_L = \left( \begin{array}{c}
               \nu^a_L\\ e^a_L\\ (\nu^c_L)^a
               \end{array}  \right) \sim (1, 3, -1/3), e^a_R
\sim (1, 1, -1),
\label{l}
\end{equation}
where a = 1, 2, 3 is the generation index.

Two of the three quark generations transform identically and
one generation transforms in a different representation:
\begin{equation}
Q_{iL} = \left( \begin{array}{c}
                d_{iL}\\-u_{iL}\\ d'_{iL}\\
                \end{array}  \right) \sim (3, \bar{3}, 0),
\label{q}
\end{equation}
\[ u_{iR}\sim (3, 1, 2/3), d_{iR}\sim (3, 1, -1/3),
d'_{iR}\sim (3, 1, -1/3),\ i=1,2,\]
\[ Q_{3L} = \left( \begin{array}{c}
                 u_{3L}\\ d_{3L}\\ T_{L}
                \end{array}  \right) \sim (3, 3, 1/3),\]
\[ u_{3R}\sim (3, 1, 2/3), d_{3R}\sim (3, 1, -1/3),
T_{R}\sim (3, 1, 2/3).\]

Fermion mass generation and symmetry breaking can be
achieved with just three $SU(3)_{L}$ triplets:
\[ \chi  \sim (1, 3, -1/3),\
\rho  \sim (1, 3, 2/3),\
\eta  \sim (1, 3, -1/3),\]
with the following vacuum expectation values (VEVs):
$\langle\chi \rangle^T = (0, 0, \omega/\sqrt{2})$,\
$\langle\rho \rangle^T = (0, u/\sqrt{2}, 0)$,\
$\langle\eta \rangle^T = (v/\sqrt{2}, 0, 0)$.
In the present model the neutrinos remain massless
at the tree level, and by radiative corrections they will
gain masses~\cite{wbm}. In this
model, the exotic quarks carry electric charges 2/3 and -1/3,
respectively, similarly to ordinary quarks. Consequently,
the exotic quarks can mix with the ordinary ones. This type
of mixing gives the flavor changing neutral currents and has
been discussed in Ref.~\cite{ll}.

By identifying $\sqrt{2} W^+_{\mu} = W^1_{\mu} - i W^2_{\mu},
\sqrt{2} Y^-_{\mu} = W^6_{\mu} - i W^7_{\mu},
\sqrt{2} X^o_{\mu} = W^4_{\mu} - i W^5_{\mu},$
the interactions among the charged vector fields with leptons are
\begin{eqnarray}
{\cal L}^{CC}_l& =& - \frac{g}{\sqrt{2}}(\bar{\nu}^a_L\gamma^\mu
e^a_LW^+_\mu + \bar{(\nu^c_L)}^a\gamma^\mu e^a_LY^+_\mu \nonumber \\
& & + \bar{\nu}^a_L\gamma^\mu (\nu^c_L
)^aX^0_\mu + \mbox{h.c.}).
\label{ccl}
\end{eqnarray}
For the quarks we have
\begin{eqnarray}
{\cal L}^{CC}_q &=&- \frac{g}{\sqrt{2}}[(\bar{u}_{3L}\gamma^\mu d_{3L}+
\bar{u}_{iL}\gamma^\mu d_{iL})W^+_\mu +
(\bar{T}_{L}\gamma^\mu d_{3L}+\bar{u}_{iL}\gamma^\mu d'_{iL})Y^+_\mu
\nonumber \\
                 & & + (\bar{u}_{3L}\gamma^\mu T_{L}-\bar{d'}_{iL}
\gamma^\mu d_{iL})X^0_\mu + \mbox{h.c.}].
\label{ccq}
\end{eqnarray}
>From (3) and (4), it follows that
the interactions with the $Y^+$ and $X^0$ bosons
violate the lepton number  and the weak isospin .

The physical neutral gauge bosons are defined through
the mixing angle $\phi$ and $Z,Z'$:
\begin{eqnarray}
Z^1  &=&Z\cos\phi - Z'\sin\phi,\nonumber\\
Z^2  &=&Z\sin\phi + Z'\cos\phi,
\end{eqnarray}
where the photon field $A_\mu$ and $Z,Z'$ are given by~\cite{hnl}:
\begin{eqnarray}
A_\mu  &=& s_W  W_{\mu}^3 + c_W\left(-\frac{t_W}{\sqrt{3}}\ W^8_{\mu}
+\sqrt{1-\frac{t^2_W}{3}}\  B_{\mu}\right),\nonumber\\
Z_\mu  &=& c_W  W^3_{\mu} + s_W\left(-\frac{t_W}{\sqrt{3}}\ W^8_{\mu}+
\sqrt{1-\frac{t_W^2}{3}}\  B_{\mu}\right), \nonumber \\
Z'_\mu &=& \sqrt{1-\frac{t_W^2}{3}}\  W^8_{\mu}+
\frac{t_W}{\sqrt{3}}\ B_{\mu}.
\label{apstat}
\end{eqnarray}

The neutral current interactions can be written in the form
\begin{eqnarray}
{\cal L}^{NC}&=&\frac{g}{2c_W}\left\{\bar{f}\gamma^{\mu}
[a_{1L}(f)(1-\gamma_5) + a_{1R}(f)(1+\gamma_5)]f
Z^1_{\mu}\right.\nonumber\\
             & & + \left.\bar{f}\gamma^{\mu}
[a_{2L}(f)(1-\gamma_5) + a_{2R}(f)(1+\gamma_5)]f
Z^2_{\mu}\right\}.
\label{nc}
\end{eqnarray}
Using $\bar{\nu}^c_L\gamma^\mu\nu^c_L=-\bar{\nu}_R
\gamma^\mu\nu_R$ we see that in this model the neutrinos have
both left- and right-handed neutral currents:
\begin{eqnarray}
a_{1L}(\nu)&=&\frac{1}{2}\left(\cos\phi +\frac{1-2s_W^2}{
\sqrt{3-4s_W^2}}\sin\phi\right),\
a_{1R}(\nu)=\frac{c^2_W}{\sqrt{3-4s_W^2}}\ \sin\phi,\nonumber\\
a_{2L}(\nu)&=&\frac{1}{2}\left(\sin\phi-
\frac{1-2s_W^2}{
\sqrt{3-4s_W^2}}\cos\phi\right), \
a_{2R}(\nu)=-\frac{c^2_W}{\sqrt{3-4s_W^2}}\ \cos\phi.
\label{van}
\end{eqnarray}
>From the above formulas, we see that coupling of the RH neutrino
neutral current with $Z^1$ is
very weak because of its linear dependence  on $\sin \phi$.

The data from the $Z$-decay allow us to
fix  the limit for $\phi$ as~\cite{hepph}  $-0.00285 \leq \phi
\leq 0.00018$. As in Ref.~\cite{hnl}, with this mixing angle,
$R_b$ in this model still disagrees with the recent experimental
value $R_b=0.2192\pm 0.0018$ measured at LEP.

To get constraint on  the new neutral gauge boson mass $M_{Z^2}$
we consider neutrino-electron scatterings.
Since in this model neutrinos have both
left- and right-handed currents, the effective four-fermion
interactions relevant to $\nu$-fermion neutral current processes,
are presented as:
\begin{eqnarray}
-{\cal L}^{\nu f}_{eff}&=&\frac{2 \rho_1 G_F}{\sqrt{2}}\left
\{g_{1V}(\nu) \bar{\nu}\gamma_\mu (1-r\gamma_5)\nu \bar{f}
\gamma^\mu[g_{1V}(f)-g_{1A}(f)\gamma_5]f\right.\nonumber\\
                       & & + \left.\xi g_{2V}(\nu)\bar{\nu}
\gamma_\mu (1-r'\gamma_5)\nu \bar{f}\gamma^\mu[g_{2V}(f)
-g_{2A}(f)\gamma_5]f\right\},
\end{eqnarray}
where  $\xi=\frac{M_{Z_1}^2}{M_{Z_2}^2}$, and
$r=\frac{g_{1A}(\nu)}{g_{1V}(\nu)},
r'=\frac{g_{2A}(\nu)}{g_{2V}(\nu)} $ are {\it right-handedness}
of currents.

The Feynman amplitude for the $\nu_\mu - e$ scattering is
\begin{eqnarray}
T_{if}&=&\frac{2 \rho_1 G_F}{\sqrt{2}}\left\{\bar{\nu}(k')
\gamma_\mu (1-r\gamma_5)\nu (k) \bar{e}(p')\gamma^\mu[g_{1V}
(\nu)g_{1V}(e) - g_{1V}(\nu)g_{1A}(e)\gamma_5]e(p)\right.
\nonumber\\
                       & & + \left.\xi\bar{\nu} (k')\gamma_\mu
(1-r'\gamma_5)\nu (k) \bar{e} (p')\gamma^\mu[ g_{2V}(\nu)
g_{2V}(e) - g_{2V}(\nu) g_{2A}(e)\gamma_5]e(p)\right\}.
\end{eqnarray}
As in Ref.~\cite{cl} in the laboratory reference frame
($\vec{p}_e =0$), the cross section is given:
\begin{equation}
\frac{d\sigma(\nu_\mu e)}{dE_e}=\frac{1}{32\pi m_e E^2_\nu}
\left(\frac{1}{2.s}\sum |M|^2\right),
\end{equation}
where $E_{\nu},E_e$ are the initial neutrino and final
electron energies and s is the number of the neutrino states.
Perform the usual manipulations~\cite{cl}, we get finally
\begin{eqnarray}
\sigma(\nu_\mu e)&=&\frac{\rho^2_1 m_e E_\nu G_F^2}{s\pi}
\left[(I^e + J^e) + \frac{1}{3}(I^e-J^e)
\right],\nonumber\\
\sigma(\bar{\nu}_\mu e)&=&\frac{\rho^2_1 m_e E_\nu G_F^2}{s\pi}
\left[\frac{1}{3}(I^e+J^e)+(I^e-J^e)\right],
\label{gf}
\end{eqnarray}
where
\begin{eqnarray}
I^e&=&g^2_{1V}(\nu)[g_{1V}^2(e)+g_{1A}^2(e)](1+r^2)+2\xi
g_{1V}(\nu) g_{2V}(\nu)
[g_{1V}(e)g_{2V}(e) + g_{1A}(e)g_{2A}(e)]\nonumber\\
 & & \times (1 + r r') +
\xi^2 g^2_{2V}(\nu)[g^2_{2V}(e) + g^2_{2A}(e)](1+r^{,2}),
\nonumber\\
J^e&=&4 rg^2_{1V}(\nu)g_{1V}(e)g_{1A}(e) + 2\xi (r + r')
g_{1V}(\nu)g_{2V}(\nu)[g_{1V}(e)g_{2A}(e)+g_{1A}(e)g_{2V}(e)]
\nonumber\\
 & & + 4 \xi^2r'g^2_{2V}(\nu)g_{2V}(e)g_{2A}(e).
\end{eqnarray}
It is easy to see that for $ r=1$ and
$\xi=0, I^e$ and $J^e$ become, respectively,
\begin{eqnarray}
I^e&=&2 g^2_{1V}(\nu)[g_{1V}^2(e)+g_{1A}^2(e)],\nonumber\\
J^e&=&4 g^2_{1V}(\nu)g_{1V}(e)g_{1A}(e).
\end{eqnarray}
Thus, it is straightforward to see that if s=1 the cross
sections in (12)  get the SM forms~\cite{cl}.

Substituting coupling constants into Eqs.(~\ref{gf}),
we finally get
\begin{eqnarray}
\sigma(\nu_\mu e)&=&\frac{\rho^2_1 m_e E_\nu G_F^2}{s 6\pi}\left(
\cos2\phi+\frac{1-2s_W^2}{\sqrt{3-4s_W^2}}\sin2\phi\right)^2
\nonumber\\
                 & & \times \left\{(1-4s_W^2+8s_W^4)[(1-\xi)^2+
(r - \xi r')^2]\right.\nonumber\\
                 & & + \left.(1-4s_W^2)(1-\xi)(r - \xi r')
\right\},\\
\sigma(\bar{\nu}_\mu e)&=&\frac{\rho^2_1 m_e E_\nu G_F^2}{s 6\pi}
\left(\cos2\phi+\frac{1-2s_W^2}{\sqrt{3-4s_W^2}}\sin2\phi\right)^2
\nonumber\\
                 & & \times \left\{(1-4s_W^2+8s_W^4)[(1-\xi)^2+
(r - \xi r')^2]\right.\nonumber\\
                 & & - \left.(1-4s_W^2)(1-\xi)(r - \xi r')\right\}.
\label{td}
\end{eqnarray}
>From Eqs.(15, 16), we see that when $\xi=0$, $\phi = 0$,
and $ r = r'= 1$, the low-energy SM results are recovered if
and only if $s=1$. Note that the formulas (15, 16) are
applicable for the theories with one extra neutral gauge
boson $Z^2$ and neutrinos having left- and right-handed components.

For our model, $r, r'$ have the forms:
\begin{eqnarray}
r(\nu)&\simeq&1-\frac{4c_W^2}{\sqrt{3-4s_W^2}}\tan\phi +
O(\tan^2\phi), \nonumber\\
r'(\nu)&\simeq&-\frac{1}{3-4s_W^2}\left(1+
\frac{4c^2_W}{\sqrt{3-4s_W^2}}\tan\phi\right) + O(\tan^2\phi).
\label{rrph}
\end{eqnarray}
Thus, Eqs.(15, 16) have only two variables $\phi$ and $\xi$. Moreover,
$\phi$ has been determined by the $Z$ decay data, consequently
$\xi$ can be fixed by the neutrino- electron scatterings.
A fit to experimental data~\cite{ah} gives a limit for the mass
of the $Z^2$ boson $M_{Z^2}\geq 320$ GeV.

We have presented the 331 model with neutrino right-handed currents.
The main features of this model may be summarized as follows:
Fermion mass generation and symmetry breaking can be
achieved with just three Higgs triplets.
The neutrinos, however,  remain massless at the tree level.
The lepton number and the weak isospin are violated in both
the heavy charged gauge boson sector and in the Higgs sector.
We have shown that in order to get the low energy SM result,
right-handed component of neutrinos in this model has to
be considered as a correction instead of an equivalent
spin state (spin-average factors of $\frac{1}{2}$).

Because of the $Z-Z'$ mixing, there is
a modification to the $Z^1$ coupling proportional to $\sin\phi$,
and the $Z$-decay gives  $-0.00285\leq \phi \leq 0.00018$.
The data from neutrino neutral current elastic scatterings shows
that mass of the new neutral gauge boson  $M_{Z^2}$ is in the
range of 400 GeV, and from the symmetry-breaking hierarchy we get:
$M_{Y^+}\simeq M_{X^o}\simeq 0.72 M_{Z^2} \geq 290$ GeV.

As the other 331 versions, the model is only anomaly free if
the number of generations is a multiple of three, and the
third generation has been treated differently from the first two.
With many unique proporties,
this model deserves our attention and further study.

\end{document}